\documentclass[onecolumn,showpacs,preprint,preprintnumbers,amsmath,amssymb]{revtex4}
\usepackage{amssymb}
\usepackage[dvips]{graphicx}
\begin{document}

\title{ Superlight small bipolarons: a route to room temperature superconductivity}
\author{A. S. Alexandrov}

\affiliation{Department of Physics, Loughborough University,
Loughborough LE11 3TU, United Kingdom\\}

\begin{abstract}
Extending the BCS theory towards the strong electron-phonon
interaction (EPI), a charged Bose liquid of small bipolarons has
been predicted by us  with a further prediction that the highest
superconducting critical temperature is found in the crossover
region of the EPI strength from the BCS-like to bipolaronic
superconductivity. Later on we have shown that the unscreened
(infinite-range) Fr\"ohlich EPI combined with the strong Coulomb
repulsion create \emph{superlight} small bipolarons, which are
several orders of magnitude  lighter than  small bipolarons in the
Holstein-Hubbard model (HHM) with a zero-range EPI.  The analytical
and numerical studies of this Coulomb-Fr\"ohlich model (CFM) provide
the following recipes for room-temperature superconductivity: (a)
The parent compound should be an ionic insulator with light ions to
form high-frequency optical phonons, (b) The structure should be
quasi two-dimensional to ensure poor screening of high-frequency
phonons polarized perpendicular to the conducting planes, (c) A
triangular lattice is required in combination with strong, on-site
 Coulomb repulsion to form the small superlight
bipolaron,  (d) Moderate carrier densities are required to keep the
system of small bipolarons close to the  Bose-Einstein condensation
regime. Clearly most of these conditions are already met in the
cuprates.

\end{abstract}


\maketitle

\section{Introduction}
The discovery of high-temperature superconductivity in cuprates
\cite{bed} has widened significantly our horizons of the theoretical
understanding of the physical phenomenon. A great number of
observations point to the possibility that the cuprate
superconductors may not be conventional Bardeen-Cooper-Schrieffer
(BCS) superconductors \cite{bcs}, but rather derive from the
Bose-Einstein condensation (BEC) of real-space small bipolarons
\cite{alemot,alebook,edwards}. Importantly a first proposal for high
temperature superconductivity  by Ogg Jr in 1946 \cite{ogg}, already
involved  real-space pairing of individual electrons into bosonic
molecules  with zero total spin. This idea was further developed as
a natural explanation of conventional superconductivity by Schafroth
\cite{sha}  and Butler and Blatt \cite{blatt}. Unfortunately the
Ogg-Schafroth picture was practically forgotten because it neither
accounted quantitatively for the critical behavior of conventional
superconductors, nor did it explain the microscopic nature of
attractive forces which could overcome the Coulomb repulsion between
two electrons  constituting a
 pair. On the contrary highly successful for low-$T_c$ metals and alloys
the BCS theory, where two electrons  were indeed  correlated, but at
a very large distance of about $10^{3}$ times of the average
inter-electron spacing, led many researchers to believe that any
superconductor  is a  "BCS-like".

However it has been found--- unexpectedly
 for many
 researchers--- that   the BCS theory and its extension \cite{eli} towards the intermediate coupling regime, $\lambda \lesssim
 1$,
   break  down already at $\lambda \gtrsim 1$ \cite{ale0}.
It happens since the Migdal "noncrossing"
 approximation \cite{mig} of the theory is not applied at $\lambda \gtrsim1$. In fact, the small parameter of the theory,
$\lambda\omega_0/E_F$, becomes large at $\lambda \gtrsim 1$ because
the bandwidth is narrowed and the Fermi energy, $E_F$  is
renormalised down exponentially due to the small polaron formation
\cite{ale0,alexandrov:2001} (here $\omega_0$ is the characteristic
phonon frequency, and we take $\hbar=c=k_B=1$). Extending the BCS
theory towards the strong interaction between electrons and ion
vibrations, $\lambda \gg 1$, a charged Bose gas of tightly bound
small bipolarons was predicted
 \cite{aleran} instead of Cooper pairs, with a further prediction  that the highest superconducting transition
temperature is attained in the crossover region  of EPI strength,
$\lambda \thickapprox 1$,  between the BCS  and  bipolaronic
superconductivity \cite{ale0}.

For a very strong EPI polarons become self-trapped on a single
lattice site and bipolarons are on-site singlets.  A finite on-site
bipolaron mass appears only in the second order of polaron
 hopping, \cite{aleran},
 so that on-site bipolarons might be very heavy in HHM, where EPI is   short-ranged.
 Actually
 HHM led some authors to
 the conclusion that the formation
of itinerant small polarons and bipolarons in real materials is
unlikely \cite{mel}, and high-temperature bipolaronic
superconductivity is impossible \cite{and2}. Nevertheless treating
the onsite repulsion (Hubbard $U$) and the short-range EPI on an
equal footing led several authors to the opposite conclusion with
respect to bipolaron mobility even in HHM, which is generally
unfavorable for coherent tunnelling.   Aubry \cite{aub93} found
along with the
 onsite bipolaron ($S0$) also an anisotropic pair of polarons
lying on two neighboring sites (i.e. the $intersite$ bipolaron,
$S1$) with classical phonons in the extreme adiabatic limit. Such
bipolarons were originally hypothesized in
\cite{alexandrov:1991,alegap} to explain the anomalous nuclear
magnetic relaxation (NMR)  in cuprate superconductors. The intersite
bipolaron could take a form of a "quadrisinglet" ($QS$) in 2D HHM,
where the electron density at the central site is 1 and "1/4" on the
four nearest neigbouring sites. In a certain region of $U$, where
$QS$ is the ground state, the double-well potential barrier which
usually pins polarons and bipolarons to the lattice depresses to
almost zero, so that adiabatic lattice bipolarons can be rather
mobile.

Mobile $S1$  bipolarons were found in 1D HHM using variational
methods also in the non and near-adiabatic regimes with dynamical
quantum phonons \cite{lamagna,trugman}. The intersite bipolaron with
a relatively small effective mass is stable in a wide region of the
parameters of HHM due to both exchange and nonadiabaticity effects
\cite{lamagna}. Near the strong coupling limit the mobile $S1$
bipolaron has an effective mass  of the order of a single Holstein
polaron mass, so that one should not rule out the possibility of a
superconducting state of $S1$ bipolarons with $s$ or $d$-wave
symmetry in HHM \cite{trugman}. More recent diagrammatic Monte Carlo
study \cite{Macridin} found $S1$ bipolarons for large $U$ at
intermediate and large EPI and established the phase diagram of 2D
HHM, comprising  unbound polarons, $S0$ and $S1$ domains. Ref.
\cite{Macridin} emphasised that the transition to the bound state
and the properties of the bipolaron in HHM are very different from
bound states in the attractive (negative $U$) Hubbard model without
EPI \cite{micnas1981}.

In any case the Holstein model is an extreme polaron model, with
typically
 highest possible values of the (bi)polaron mass in the strong
coupling regime \cite{ale5,alekor,spencer,jim}. Many doped ionic
lattices, including cuprates, are characterized by poor screening of
high-frequency optical phonons and they are more appropriately
described by the finite-range Fr\"ohlich EPI. The unscreened
Fr\"ohlich EPI provides relatively light lattice polarons
 and combined with the Coulomb repulsion  also "superlight" but yet small (intersite) bipolarons.  In contrast with the crawler motion of
on-site bipolarons, the intersite-bipolaron tunnelling
 is a crab-like, so that the effective mass
scales linearly with the polaron mass. Such bipolarons are several
orders of magnitude lighter than small bipolarons in HHM
\cite{ale5}. Here I review a few analytical \cite{ale5, alekor2} and
more recent Quantum Monte-Carlo (QMC)  \cite{jim2,jim3} studies of
CFM which  have found superlight bipolarons  in a wide parameter
range with achievable phonon frequencies and couplings. They  could
 have a superconducting transition in excess of room
temperature.

\section{Coulomb-Fr\"ohlich model}

Any realistic theory of doped narrow-band ionic insulators should
include both the finite-range Coulomb repulsion and the strong
finite-range EPI. From a theoretical standpoint, the inclusion of
the finite-range Coulomb repulsion is critical in ensuring that the
carriers would not form clusters.   The  Coulomb repulsion, $V_{c}$,
makes the clusters unstable and lattice bipolarons \emph{more
mobile}.

To illustrate the point let us consider a generic multi-polaron
"Coulomb-Fr\"{o}hlich" model (CFM) on a lattice,
 which explicitly includes the
finite-range Coulomb repulsion, $V_c$, and the strong long-range EPI
\cite{ale5,alekor2}. The implicitly present (infinite)
 Hubbard $U$ prohibits double occupancy and removes the
need to distinguish the fermionic spin, if we are interested in the
charge rather than spin excitations. Introducing spinless fermion
operators $c_{{\bf n}}$ and phonon operators $d_{{\bf m}}$, the
Hamiltonian of CFM is written in the real-space representation
 as \cite{alekor2}
\begin{eqnarray}
H &=&\sum_{{\bf n\neq n^{\prime }}}T({\bf n-n^{\prime }})c_{{\bf n}%
}^{\dagger }c_{{\bf n^{\prime }}}+{1\over {2}}\sum_{{\bf n\neq n^{\prime }}}V_{c}({\bf %
n-n^{\prime }})c_{{\bf n}}^{\dagger }c_{{\bf n}}c_{{\bf n^{\prime }}%
}^{\dagger }c_{{\bf n^{\prime }}}+ \\
&&\omega _{0}\sum_{{\bf n\neq m} }g({\bf m-n})({\bf e}\cdot {\bf
e}_{{\bf m-n}})c_{{\bf n}}^{\dagger }c_{{\bf n}}(d_{{\bf m}
}^{\dagger }+d_{{\bf m} })+  \nonumber \\
&&\omega _{0}\sum_{{\bf m}}\left( d_{{\bf m} }^{\dagger }d_{{\bf m}%
}+\frac{1}{2}\right),  \nonumber
\end{eqnarray}
where $T({\bf n})$ is the bare hopping integral in a rigid lattice.
In general, this many-body model is of considerable complexity.
However, if we are interested in the non or near adiabatic limit and
the strong EPI,  the kinetic energy is a perturbation.  Then the
model can be grossly simplified using the Lang-Firsov canonical
transformation \cite{lan} in the Wannier representation for
electrons and phonons,
\[
S=\sum_{{\bf m\neq n} }g({\bf m-n})({\bf e}\cdot {\bf e}_{%
{\bf m-n}})c_{{\bf n}}^{\dagger }c_{{\bf n}}(d_{{\bf m}}^{\dagger }-d_{%
{\bf m}}).
\]

Here we consider a particular lattice structure, where intersite
lattice bipolarons tunnel already in the first order in $T({\bf
n})$. That allows us to average the transformed Hamiltonian,
$\tilde{H}=\exp(S) H \exp(-S)$ over phonons to obtain
\begin{equation}
\tilde{H}=H_{0}+H_{pert}, \label{trans}
\end{equation}
where
\[
H_{0}=-E_{p}\sum_{{\bf n}}c_{{\bf n}}^{\dagger }c_{{\bf n}}+{1\over
{2}}\sum_{{\bf n\neq
n^{\prime }}}v({\bf n-n^{\prime }})c_{{\bf n}}^{\dagger }c_{{\bf n}}c_{{\bf %
n^{\prime }}}^{\dagger }c_{{\bf n^{\prime }}} +\omega _{0}\sum_{{\bf m}}\left( d_{{\bf m} }^{\dagger }d_{{\bf m}%
}+\frac{1}{2}\right),
\]
and
\[
H_{pert}=\sum_{{\bf n\neq n^{\prime }}}t({\bf n-n^{\prime }})c_{{\bf n}%
}^{\dagger }c_{{\bf n^{\prime }}}.
\]
is a perturbation. $E_{p}$ is the familiar polaron level shift,
\begin{equation}
E_{p}=\omega_0 \sum_{{\bf m}\nu }g^{2}({\bf m-n})({\bf e}\cdot {\bf
e}_{{\bf m-n}})^{2}, \label{polshift}
\end{equation}
which is independent of ${\bf n}$. The polaron-polaron interaction
is
\begin{equation}
v({\bf n-n^{\prime }})=V_{c}({\bf n-n^{\prime }})-V_{ph}({\bf n-n^{\prime }}%
),
\end{equation}
where
\begin{equation}
V_{ph}({\bf n-n^{\prime }}) =2\omega _{0}\sum_{\bf m}g({\bf %
m-n})g({\bf m-n^{\prime }})
({\bf e}\cdot {\bf e}_{{\bf m-n}})({\bf e}\cdot {\bf e}_{%
{\bf m-n^{\prime }}}). \label{v}
\end{equation}
The transformed  hopping integral  is $t({\bf n-n^{\prime }})=T({\bf
n-n^{\prime }})\exp [-g^{2}({\bf n-n^{\prime }})]$ with
\begin{eqnarray}
g^{2}({\bf n-n^{\prime }}) &=&\sum_{{\bf m},\nu }g({\bf m-n})({\bf e}\cdot {\bf e}_{{\bf m-n}})\times \\
&&\left[ g({\bf m-n})({\bf e}\cdot {\bf e}_{{\bf m-n}})-g({\bf m-n^{\prime }})({\bf e}\cdot {\bf e}_{{\bf m-n^{\prime }}})%
\right]  \nonumber
\end{eqnarray}
at  low temperatures. The mass renormalization exponent can be
expressed via $E_{p}$ and $V_{ph}$ as
\begin{equation}
g^{2}({\bf n-n^{\prime }})=\frac{1}{\omega _{0}}\left[ E_{p}-\frac{1}{2}%
V_{ph}({\bf n-n^{\prime }})\right] . \label{g2}
\end{equation}

The Hamiltonian $\tilde{H}$, Eq.(\ref{trans}), in zero order with
respect to the hopping describes localised polarons and independent
phonons, which are vibrations of ions relative to new equilibrium
positions depending on the polaron occupation numbers. Importantly
the phonon frequencies remain unchanged in this limit at any polaron
density, $n$. At finite $\lambda$ and $n$ there is a softening of
phonons $\delta \omega_0$ of the order of $\omega_0 n/\lambda^2$
\cite{ale2}. Interestingly the optical phonon can be mixed with a
low-frequency polaronic plasmon forming a new excitation,
"plasphon", which was proposed in \cite{plasphon,ale2} as an
explanation of the anomalous phonon mode splitting observed in
cuprates \cite{splitting}. The middle of the electron band is
shifted down by the polaron level-shift $E_{p}$ due to the potential
well created by lattice deformation.

When $V_{ph}$ exceeds $V_{c}$ the full interaction becomes negative
and polarons form pairs. The real space representation allows us to
elaborate more physics behind the lattice sums in $V_{ph}$
\cite{alekor2}. When a carrier (electron or hole) acts on an ion
with a force ${\bf f}$, it displaces the ion by some vector ${\bf
x}={\bf f}/k$. Here $k$ is the ion's
force constant. The total energy of the carrier-ion pair is $-{\bf f}%
^{2}/(2k)$. This is precisely the summand in Eq.(\ref{polshift})
expressed via dimensionless coupling constants. Now consider two
carriers interacting with
the {\em same} ion. The ion displacement is ${\bf x}=({\bf f}%
_{1}+{\bf f}_{2})/k$ and the energy is $-{\bf f}_{1}^{2}/(2k)-{\bf f}%
_{2}^{2}/(2k)-({\bf f}_{1}\cdot {\bf f}_{2})/k$. Here the last term
should be interpreted as an ion-mediated interaction between the two
carriers. It depends on the scalar product of ${\bf f}_{1}$ and
${\bf f}_{2}$ and consequently on the relative positions of the
carriers with respect to the ion. If the ion is an isotropic
harmonic oscillator,  then the following simple rule applies. If the
angle $\phi $ between ${\bf f}_{1}$ and ${\bf f}_{2}$ is less than
$\pi /2$ the polaron-polaron interaction will be attractive, if
otherwise it will be repulsive. In general, some ions will generate
attraction, and some repulsion between polarons.

The overall sign and magnitude of the interaction is given by the
lattice sum in Eq.(\ref{v}). One should  note that according to
Eq.(\ref{g2}) an attractive EPI reduces the polaron mass (and
consequently the bipolaron mass), while repulsive EPI enhances the
mass. Thus, the long-range EPI serves a double purpose. Firstly, it
generates an additional inter-polaron attraction because the distant
ions have small angle $\phi $. This additional attraction helps to
overcome the direct  Coulomb repulsion between
 polarons. And secondly, the Fr\"{o}hlich EPI makes lattice
(bi)polarons lighter. Here, following \cite{ale5,alekor2,jim2,jim3},
we consider a few examples of  intersite superlight bipolarons.

\section{Apex bipolarons}
High-$T_{c}$ oxides  are doped charged-transfer ionic insulators
with narrow electron bands. Therefore, the interaction between holes
can be analyzed using computer simulation techniques based on a
minimization of the ground state energy of an ionic insulator with
two holes, the lattice deformations and the Coulomb repulsion fully
taken into account, but neglecting the kinetic energy terms. Using
these techniques net inter-site interactions of the in-plane oxygen
hole with the $apex$ hole, Fig.1, and of two in-plane oxygen holes,
Fig.2, were found to be attractive in $La_{2}CuO_{4}$ \cite{cat}
with the binding energies $\Delta =119meV$ and $\Delta =60meV$,
respectively.  All other interactions were found to be repulsive.

Both apex and in-plane bipolarons can tunnel from one
unit cell to another via the {\it %
\ single}-polaron tunnelling from one apex  oxygen to its apex
neighbor in case of the apex bipolaron \cite{ale5}, Fig.1, or via
the next-neighbor hopping in case of the in-plane bipolaron
\cite{alekor2}, Fig.2.
\begin{figure}[tbp]
\begin{center}
\includegraphics[angle=-90,width=0.47\textwidth]{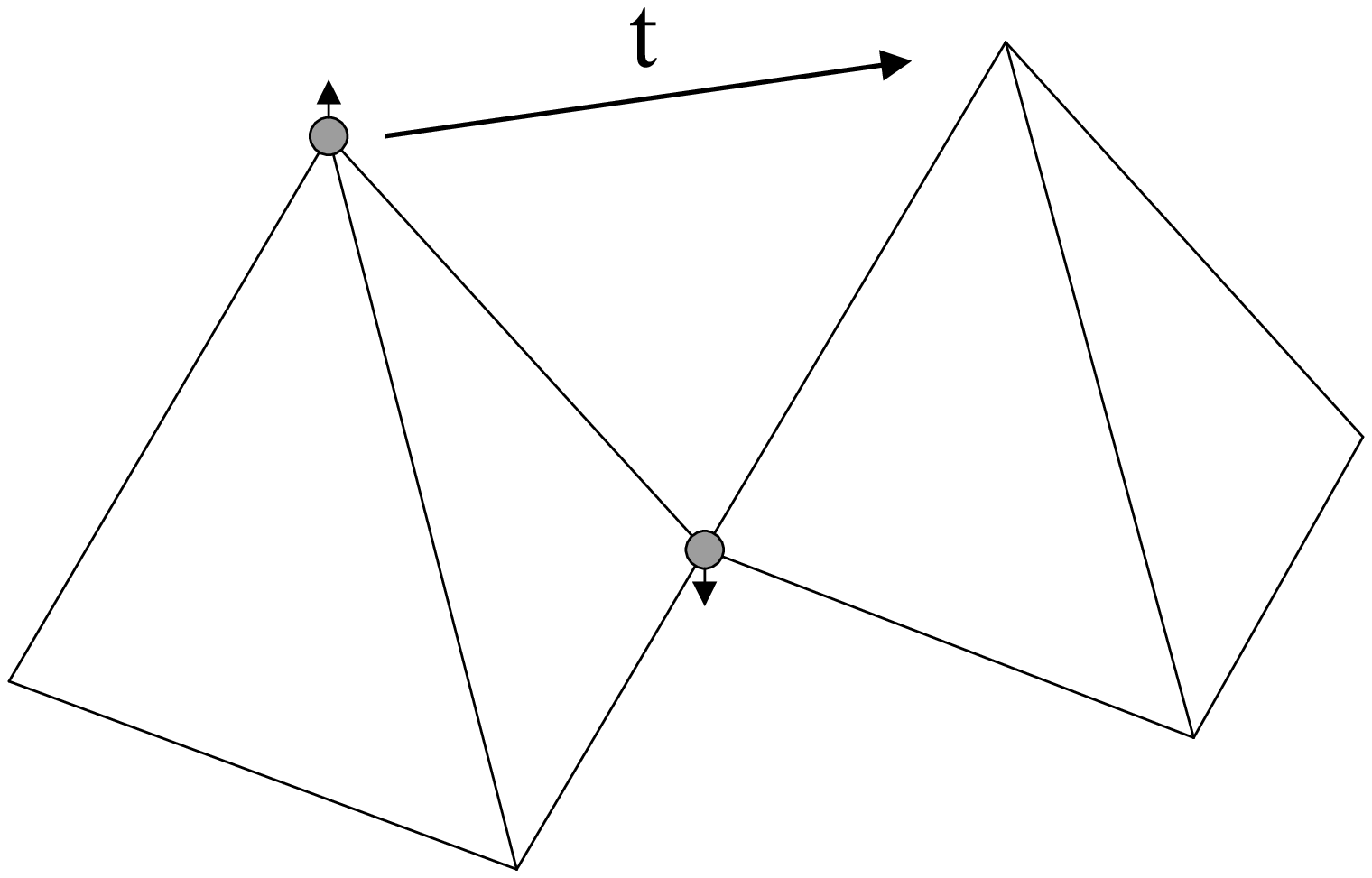} \vskip -0.5mm
\end{center}
\caption{Apex bipolaron tunnelling in perovskites (after
\cite{ale5})}
\end{figure}
The Bloch bands of these bipolarons  are obtained using the
canonical transformation, described above,  projecting the
transformed Hamiltonian, Eq.(\ref{trans}), onto a reduced Hilbert
space containing only empty or doubly occupied elementary cells
\cite{alebook}. The wave function of the apex bipolaron localized,
 say in cell ${\bf m}$ is written as
\begin{equation}
|{\bf m}\rangle =\sum_{i=1}^{4}A_{i}c_{i}^{\dagger
}c_{apex}^{\dagger }|0\rangle ,
\end{equation}
where $i$ denotes the $p_{x,y}$ orbitals and spins of the four plane
oxygen ions in the cell, Fig.1, and $c_{apex}^{\dagger }$ is the
creation operator for the hole in one of the three apex oxygen
orbitals with the spin, which is the same or opposite to the spin of
the in-plane hole depending on the total spin of the bipolaron. The
probability amplitudes $A_{i}$ are
normalized by the condition $|A_{i}|=1/2,$ if four plane orbitals $%
p_{x1},p_{y2},p_{x3}$ and $p_{y4}$ are involved, or by
$|A_{i}|=1/\sqrt{2}$ if only two of them are relevant. Then a matrix
element of the Hamiltonian Eq.(\ref{trans})  describing the
\emph{bipolaron } tunnelling to the nearest neighbor cell ${\bf
m+a}$ is found as
\begin{equation}
t_b=\langle {\bf m}|\tilde{H}|{\bf m}+{\bf a}\rangle
=|A_{i}|^{2}T_{pp^{\prime }}^{apex}e^{-g^{2}},
\end{equation}
where $T_{pp^{\prime }}^{apex}e^{-g^{2}}$ is a \emph{single polaron}
hopping integral between two apex ions. The inter-site bipolaron
tunnelling appears already in \emph{the first order} with respect to
the single-hole transfer  $T_{pp^{\prime }}^{apex}$, and the
bipolaron energy
spectrum consists of two subbands $E^{x,y}({\bf K),}$ formed by the overlap of $%
p_{x}$ and $p_{y}$ $apex$ oxygen  orbitals, respectively (here we
take the lattice constant $a=1$):
\begin{eqnarray}
E^{x}({\bf K}) &=&t\cos (K_{x})-t^{\prime }\cos (K_{y}), \\
E^{y}({\bf K}) &=&-t^{\prime }\cos (K_{x})+t\cos (K_{y}). \nonumber
\end{eqnarray}

\begin{figure}[tbp]
\begin{center}
\includegraphics[angle=-0,width=0.50\textwidth]{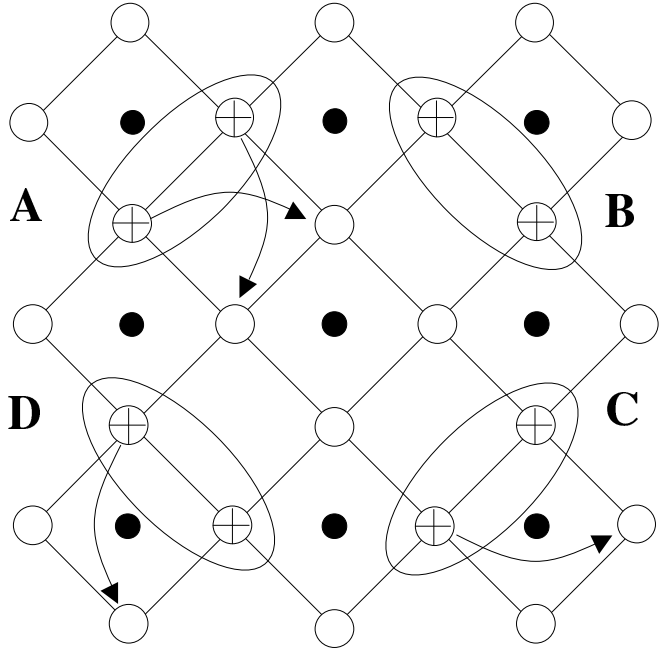} \vskip -0.5mm
\end{center}
\caption{Four degenerate in-plane bipolaron configurations A, B, C,
and D . Some single-polaron hoppings are indicated by arrows
(Reproduced from A. S. Alexandrov and P. E.  Kornilovitch,  J.
Phys.: Condens. Matter {\bf 14}, 5337 (2002), (c) IOP Publishing
Limited, 2007. \cite{alekor2}). }
\end{figure}

 They transform into one another under $\pi /2$ rotation. If
$t,t^{\prime
}>0, $ ``$x"$ bipolaron band has its minima at ${\bf K=}(\pm \pi ,0)$ and $y$%
-band at ${\bf K}=(0,\pm \pi )$. In these equations $t$ is the
renormalized hopping integral between $p$ orbitals of the same
symmetry elongated in the direction of the hopping ($pp\sigma $) and
$t^{\prime }$ is the renormalized hopping integral in the
perpendicular direction ($pp\pi $). Their ratio $t/t^{\prime
}=T_{pp^{\prime }}^{apex}/T^{\prime }{}_{pp^{\prime }}^{apex}=4$ as
follows from the tables of hopping integrals in solids. Two
different bands are not mixed because $T_{p_{x},p_{y}^{\prime
}}^{apex}=0$ for the nearest neighbors. A random potential does not
mix them either, if it varies smoothly on the lattice scale. Hence,
we can distinguish `$x$' and `$y$' bipolarons with a lighter
effective mass in $x$ or $y$ direction, respectively. The apex $z$
bipolaron, if formed, is $ca.$ four times less mobile than $x$ and
$y$ bipolarons. The bipolaron bandwidth is \emph{of the same order}
as the polaron one, which is a specific feature of  the inter-site
bipolaron. For a large part of the Brillouin zone near $%
(0,\pi )$ for `$x$' and $(\pi ,0)$ for `$y$' bipolarons, one can
adopt the effective mass approximation
\begin{equation}
E^{x,y}({\bf K})={\frac{K_{x}^{2}}{{2m_{x,y}^{\ast \ast }}}}+{\frac{%
K_{y}^{2}}{{2m_{y,x}^{\ast \ast }}}}
\end{equation}
with $K_{x,y}$ taken relative to the band bottom positions and
$m_{x}^{\ast \ast }=1/t$, $m_{y}^{\ast \ast }=4m_{x}^{\ast \ast }$.

$X$ and $y$ bipolarons  bose-condense at the boundaries of the
center-of-mass Brillouin zone with $K=(\pm\pi,0)$ and
$K=(0,\pm\pi)$, respectively, which explains the d-wave symmetry and
the checkerboard modulations of the order parameter in cuprates
\cite{aledwave}.

\section{In-plane bipolarons}
Now let us  consider  in-plane
bipolarons in a two-dimensional lattice of ideal octahedra that can
be regarded as a simplified model of the copper-oxygen perovskite
layer, Fig.3 \cite{alekor2}. The lattice period is $a=1$ and the
distance between the apical sites and the central plane is
$h=a/2=0.5$. For mathematical transparency we assume that all
in-plane atoms, both copper and oxygen, are static but apex oxygens
are independent three-dimensional isotropic harmonic oscillators.

\begin{figure}[tbp]
\begin{center}
\includegraphics[angle=-0,width=0.47\textwidth]{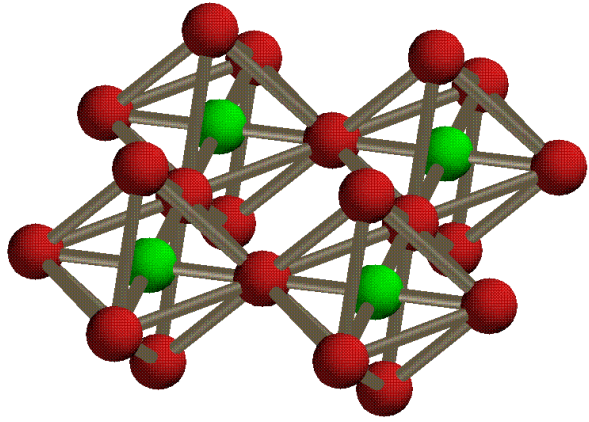} \vskip -0.5mm
\end{center}
\caption{Simplified model of the copper-oxygen perovskite layer.
Reproduced from A. S. Alexandrov and P. E.  Kornilovitch,  J. Phys.:
Condens. Matter {\bf 14}, 5337 (2002), (c) IOP Publishing Limited,
2007. \cite{alekor2} }
\end{figure}

Due to poor screening, the hole-apex interaction is purely
coulombic,
\[
g_{\alpha }({\bf m-n})=\frac{\kappa _{\alpha }}{|{\bf m-n}|^{2}},
\]
where $\alpha =x,y,z$. To account for the  fact that $c$
axis-polarized
phonons couple to the holes stronger than others due to a poor screening \cite{ale5,alekor,alekor2}, we choose $%
\kappa _{x}=\kappa _{y}=\kappa _{z}/\sqrt{2}$. The direct hole-hole
repulsion is
\[
V_{c}({\bf n-n^{\prime }})=\frac{V_{c}}{\sqrt{2}|{\bf n-n^{\prime
}}|}
\]
so that the repulsion between two holes in the nearest neighbor
($NN$) configuration is $V_{c}$. We also include the bare $NN$
hopping $T_{NN}$, the next nearest neighbor ($NNN$) hopping across
copper $T_{NNN}$ and the $NNN$ hopping between the pyramids
$T_{NNN}^{\prime }$.

The polaron shift is given by the lattice sum Eq.(3), which after
summation over polarizations yields
\begin{equation}
E_{p}=2\kappa _{x}^{2}\omega _{0}\sum_{{\bf m}}\left( \frac{1}{|{\bf m-n}%
|^{4}}+\frac{h^{2}}{|{\bf m-n}|^{6}}\right) =31.15\kappa
_{x}^{2}\omega _{0}, \label{televen}
\end{equation}
where the factor $2$ accounts for two layers of apical sites. For
reference,
the Cartesian coordinates are ${\bf n}=(n_{x}+1/2,n_{y}+1/2,0)$, ${\bf m}%
=(m_{x},m_{y},h)$, and $n_{x},n_{y},m_{x},m_{y}$ are integers. The
polaron-polaron attraction is
\begin{equation}
V_{ph}({\bf n-n^{\prime }})=4\omega \kappa _{x}^{2}\sum_{{\bf m}}\frac{%
h^{2}+({\bf m-n^{\prime }})\cdot ({\bf m-n})}{|{\bf m-n^{\prime }}|^{3}|{\bf %
m-n}|^{3}}.  \label{ttwelve}
\end{equation}
Performing the lattice summations for the $NN$, $NNN$, and
$NNN^{\prime}$ configurations one finds $V_{ph}=1.23\,E_{p},$
$0.80\,E_{p}$, and $0.82\,E_{p}$,
respectively. As a result, we obtain a net inter-polaron interaction as $%
v_{NN}=V_{c}-1.23\,E_{p}$, $v_{NNN}=\frac{V_{c}}{\sqrt{2}}-0.80\,E_{p}$, $%
v_{NNN}^{\prime }=\frac{V_{c}}{\sqrt{2}}-0.82\,E_{p}$, and the mass
renormalization exponents as $g_{NN}^{2}=0.38(E_{p}/\omega)$, $%
g_{NNN}^{2}=0.60(E_{p}/\omega)$ and $(g^{\prime
}{}_{NNN})^{2}=0.59(E_{p}/\omega)$.

Let us now discuss different regimes of the model. At
$V_{c}>1.23\,E_{p}$, no bipolarons are formed and the system is a
polaronic Fermi liquid. Polarons tunnel in the {\em square} lattice
with  $t=T_{NN}\exp (-0.38E_{p}/\omega)$ and  $t^{\prime
}=T_{NNN}\exp (-0.60E_{p}/\omega)$ for $NN$ and $NNN$ hoppings,
respectively. Since $g_{NNN}^{2}\approx (g_{NNN}^{\prime })^{2} $
one can neglect the difference between $NNN$ hoppings within and
between the octahedra. A single polaron spectrum is therefore
\begin{equation}
E_{1}({\bf k})=-E_{p}-2t^{\prime }[\cos k_{x}+\cos k_{y}]-4t\cos
(k_{x}/2)\cos (k_{y}/2).  \label{tfifteen}
\end{equation}
The polaron mass is $m^{\ast }=1/(t+2t^{\prime })$. Since in general $%
t>t^{\prime }$, the mass is mostly determined by the $NN$ hopping amplitude $t$%
.

If $V_{c}<1.23\,E_{p}$ then intersite $NN$ bipolarons form. The
bipolarons
tunnel in the plane via four resonating (degenerate) configurations $A$, $B$%
, $C$, and $D$, as shown in Fig.2. In the first order of the
renormalized hopping integral, one should retain only these lowest
energy configurations and discard all the processes that involve
configurations with higher energies. The result of such a projection
is the bipolaron Hamiltonian,
\begin{eqnarray}
H_{b} &=&(V_{c}-3.23\,E_{p})\sum_{{\bf l}}[A_{{\bf l}}^{\dagger }A_{{\bf l}%
}+B_{{\bf l}}^{\dagger }B_{{\bf l}}+C_{{\bf l}}^{\dagger }C_{{\bf l}}+D_{%
{\bf l}}^{\dagger }D_{{\bf l}}] \\
&&-t^{\prime }\sum_{{\bf l}}[A_{{\bf l}}^{\dagger }B_{{\bf l}}+B_{{\bf l}%
}^{\dagger }C_{{\bf l}}+C_{{\bf l}}^{\dagger }D_{{\bf l}}+D_{{\bf l}%
}^{\dagger }A_{{\bf l}}+H.c.]  \nonumber \\
&&-t^{\prime }\sum_{{\bf n}}[A_{{\bf l-x}}^{\dagger }B_{{\bf l}}+B_{{\bf l+y}%
}^{\dagger }C_{{\bf l}}+C_{{\bf l+x}}^{\dagger }D_{{\bf l}}+D_{{\bf l-y}%
}^{\dagger }A_{{\bf l}}+H.c.],  \nonumber
\end{eqnarray}
where ${\bf l}$ numbers octahedra rather than individual sites, ${\bf x}%
=(1,0)$, and ${\bf y}=(0,1)$. A Fourier transformation and
diagonalization of a $4\times 4$ matrix yields the bipolaron
spectrum:
\begin{equation}
E_{2}({\bf K})=V_{c}-3.23E_{p}\pm 2t^{\prime }[\cos (K_{x}/2)\pm
\cos (K_{y}/2)].  \label{tseventeen}
\end{equation}
There are four bipolaronic subbands combined in the band of the width $%
8t^{\prime }$. The effective mass of the lowest band is $m^{\ast
\ast }=2/t^{\prime }$. The bipolaron binding energy is $\Delta
\approx 1.23E_{p}-V_{c}.$ Inter-site bipolarons already move in the
{\em first} order of the single polaron hopping. This remarkable
property is entirely due to the strong on-site repulsion and
long-range electron-phonon interactions that leads to a non-trivial
connectivity of the lattice. This fact combines with a weak
renormalization of $t^{\prime }$ yielding a {\em superlight}
bipolaron with the mass $m^{\ast \ast }\propto \exp
(0.60\,E_{p}/\omega )$. We recall that in the Holstein model
$m^{\ast \ast }\propto \exp (2E_{p}/\omega )$ \cite{aleran}. Thus
the mass of the Fr\"{o}hlich bipolaron in the perovskite layer
scales approximately as a {\em cubic root} of that of the Holstein
bipolaron.

 At even stronger EPI, $V_{c}<1.16E_{p}$, $NNN$
bipolarons become stable. More importantly, holes can now form 3-
and 4-particle clusters. The dominance of the potential energy over
kinetic in the transformed Hamiltonian enables us to readily
investigate these many-polaron cases. Three holes placed within one
oxygen square have four degenerate states with the energy
$2(V_{c}-1.23E_{p})+V_{c}/\sqrt{2}-0.80E_{p}$. The first-order
polaron hopping processes mix the states resulting in a ground state
linear
combination with the energy $E_{3}=2.71V_{c}-3.26E_{p}-\sqrt{%
4t^{2}+t^{\prime }{}^{2}}$. It is essential that between the squares
such triads could move only in higher orders of polaron hopping. In
the first order, they are immobile. A cluster of four holes has only
one state within
a square of oxygen atoms. Its energy is $E_{4}=4(V_{c}-1.23E_{p})+2(V_{c}/%
\sqrt{2}-0.80E_{p})=5.41V_{c}-6.52E_{p}$. This cluster, as well as
all bigger ones, are also immobile in the first order of polaron
hopping. We would like to stress that at distances much larger than
the lattice constant the polaron-polaron interaction is always
repulsive, and the formation of infinite clusters, stripes or
strings is  prohibited. We conclude that at $V_{c}<1.16E_{p}$ the
system quickly becomes a charge segregated insulator.

The fact that within the window, $1.16E_{p}<V_{c}<1.23E_{p}$, there
are no three or more polaron bound states, indicates that bipolarons
repel each other. The system is effectively a charged Bose-gas,
which is a superconductor \cite{ogg,sha}. This superconducting state
requires a rather fine balance between electronic and ionic
interactions in cuprates.

\section{All-coupling lattice bipolarons}
The multi-polaron CFM model discussed above is analytically solvable
in the strong-coupling nonadibatic ($\omega_0 \gtrsim T(a))$ limit
using the Lang-Firsov transformation of the Hamiltonian, Eq.(1), and
projecting it on the inter-site pair Hilbert space
\cite{ale5,alekor2}. To extend the theory for the whole parameter
space an advanced continuous time QMC technique (CTQMC) has been
recently developed for bipolarons \cite{jim2,jim3}. Using CTQMC
refs. \cite{jim2,jim3}   simulated the CFM Hamiltonian on a
staggered triangular ladder (1D), triangular (2D) and strongly
anisotropic hexagonal (3D) lattices including triplet pairing
\cite{jim3}. On such lattices, bipolarons are found to move with a
crab like motion (Fig. 1), which is distinct from the crawler motion
found on cubic lattices \cite{aleran}. Such bipolarons are small but
very light for a wide range of electron-phonon couplings and phonon
frequencies. EPI has been modeled using the force function in the
site-representation as

\begin{equation}
H_{e-ph} =    - \sum_{\mathbf{n}\mathbf{m}\sigma}
f_{\mathbf{m}}(\mathbf{n}) c^{\dagger}_{\mathbf{n}\sigma}
c_{\mathbf{n}\sigma} \xi_{\mathbf{m}} \: . \label{eq:four}
\end{equation}

Each vibrating ion has one phonon degree of freedom $\xi_\mathbf{m}$
associated with a single atom. The sites are numbered by the indices
$\mathbf{n}$ or $\mathbf{m}$ for electrons and ions respectively.
Operators $c^{\dagger}_{\mathbf{n}\sigma}$ creates an electron on
site ${\bf n}$ with spin $\sigma$.  Coulomb repulsion
$V(\mathbf{n}-\mathbf{n}')$ has been  screened up to the first
nearest neighbors, with on site repulsion $U$ and nearest-neighbor
repulsion $V_c$. In contrast, the Fr\"ohlich interaction is assumed
to be long-range, due to unscreened interaction with c-axis
high-frequency phonons \cite{ale5}. The form of the interaction with
c-axis polarized phonons has been specified via the force
function\cite{alekor},
$f_{\mathbf{m}}(\mathbf{n})=\kappa\left[(\mathbf{m}-\mathbf{n})^2+1\right]^{-3/2}$,
where $\kappa$ is a constant. The dimensionless electron-phonon
coupling constant $\lambda$ is defined as
$\lambda=\sum_{\mathbf{m}}f^{2}_{\mathbf{m}}(0)/2M\omega^2 zT(a)$
which is the ratio of the polaron binding energy  to the kinetic
energy of the free electron $zT(a)$, and the lattice constant is
taken as $a=1$.

\begin{figure}
\includegraphics[height=105mm,angle=270]{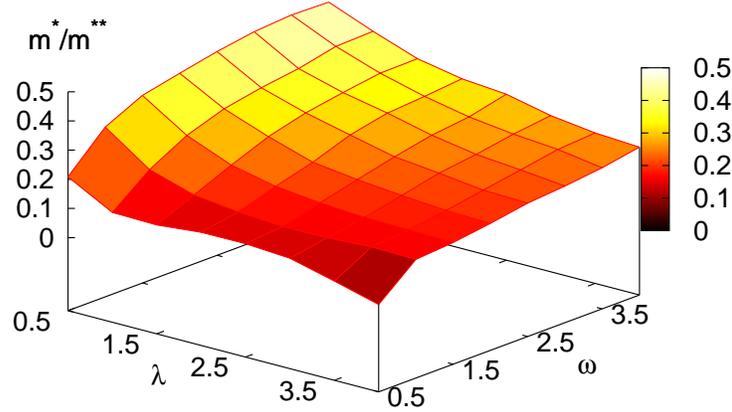}
\caption{Polaron to bipolaron mass ratio for a range of
$\bar{\omega}=\omega_0/T(a)$ and $\lambda$ on the staggered ladder.
Mobile small bipolarons are seen even in the adiabatic regime
$\bar{\omega}=0.5$ for couplings $\lambda$ up to 2.5 (Reproduced
from J. P. Hague et al., Phys. Rev. Lett {\bf 98}, 037002 (2007),
(c) American Physical Society, 2007. \cite{jim2}" ). }
\end{figure}

In the limit of high phonon frequency $\omega\gg T(a)$ and large
on-site Coulomb repulsion (Hubbard $U$), the model is reduced to an
extended Hubbard model with intersite attraction and suppressed
double-occupancy \cite{alekor2} by applying the Lang-Firsov
canonical transformation (section 2). Then the Hamiltonian can be
projected onto the subspace of nearest neighbor intersite
\emph{crab} bipolarons (sections 4). In contrast with the crawler
bipolaron, the crab bipolaron's mass scales linearly with the
polaron mass ($m^{**}=4m^*$ on the staggered chain and
$m^{**}=6m^{*}$ on the triangular lattice).

Extending the CTQMC algorithm
 to systems of two particles with
strong EPI and Coulomb repulsion solved the bipolaron problem on a
staggered ladder, triangular and anisotropic hexagonal lattices from
weak to strong coupling in a realistic parameter range where usual
strong and weak-coupling limiting approximations fail. Importantly
small but light  bipolarons have been found for more realistic
intermediate values of EPI, $\lambda \lesssim 1 $ and phonon
frequency, $\omega \lesssim T(a)$ \cite{jim2,jim3}.

\begin{figure}
\includegraphics[height=105mm,angle=270]{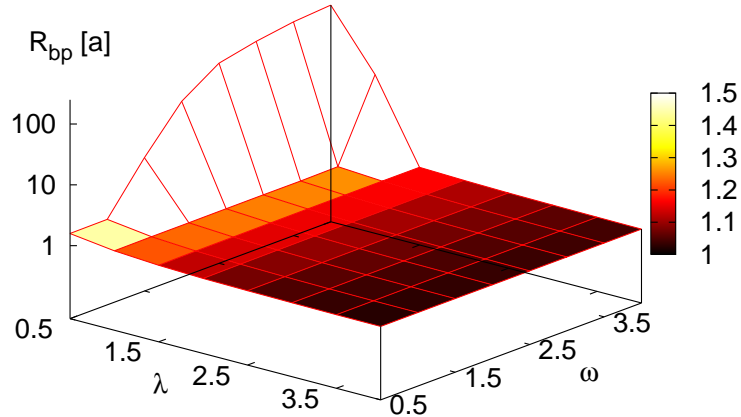}
\caption{ Bipolaron radius (in units of $a$) for a range of
$\bar{\omega}$ and $\lambda$ on the staggered ladder (Reproduced
from J. P. Hague et al., Phys. Rev. Lett {\bf 98}, 037002 (2007),
(c) American Physical Society, 2007. \cite{jim2}" ).}
\end{figure}

Figure 4 shows the ratio of the polaron to bipolaron masses on the
staggered ladder as a function of effective coupling and phonon
frequency for $V_c=0$. The bipolaron to polaron mass ratio is about
2 in the weak coupling regime ($\lambda\ll1$) as it should be for a
large bipolaron \cite{ver}. In the strong-coupling, large phonon
frequency limit the mass ratio approaches 4, in agreement with
strong-coupling arguments given above.  In a wide region of
parameter space, we find a bipolaron/polaron mass ratio of between 2
and 4 and a bipolaron radius similar to the lattice spacing, see
Figs. 4 and 5. Thus the bipolaron is small and light at the same
time. Taking into account additional intersite Coulomb repulsion
$V_c$ does not change this conclusion. The bipolaron is stable for
$V_c<4T(a)$. As $V_c$ increases the bipolaron mass decreases but the
radius remains small, at about 2 lattice spacings. Importantly, the
absolute value of the small bipolaron mass is only about 4 times of
the bare electron mass $m_0$, for $\lambda=\omega/T(a)=1$ (see Fig.
4).

Simulations of the bipolaron on an infinite triangular lattice
including exchanges and large on-site Hubbard repulsion $U = 20T(a)$
also lead to the bipolaron mass of about  $6m_{0xy}$ and the
bipolaron radius  $R_{bp}\thickapprox 2a$  for a moderate coupling
$\lambda= 0.5$ and a large phonon frequency $\omega= T(a)$ (for the
triangular lattice, $m_{0xy}=1/3a^{2}T(a)$). Finally,  the bipolaron
in a hexagonal lattice with out-of-plane hopping $T'=T(a)/3$ has
also a light in-plane inverse mass, $m_{xy}^{**}\thickapprox 4.5
m_{0xy}$ but a small size, $R_{bp}\thickapprox 2.6a$ for
experimentally achievable values of the phonon frequency
$\omega=T(a)=200$meV and EPI, $\lambda=0.36$. Out-of-plane
$m_{z}^{**}\thickapprox70m_{0z}$ is Holstein like, where
$m_{0z}=1/2d^2T'$, ($d$ is the inter-plane spacing). When bipolarons
are small and pairs do not overlap, the pairs can form  a BEC at
$T_{BEC}=3.31(2n_B/a^2\sqrt{3}d)^{2/3}/(m_{xy}^{2/3}m_z^{1/3})$. If
we choose realistic values for the lattice constants of 0.4 nm in
the plane and 0.8 nm out of the plane, and allow the density of
bosons to be $n_B$=0.12 per lattice site, which easily avoids
overlap of pairs, then $T_{BEC}\thickapprox$ 300K.

\section{Summary}

 For a very strong electron-phonon coupling in the Holstein model with  the zero-range EPI, polarons
become self-trapped on a single lattice site and bipolarons are
on-site singlets.
 The on-site  bipolaron mass  appears only in the second order of polaron hopping \cite{aleran},
 so that on-site bipolarons are very heavy. This estimate led some authors to
 the conclusion that  \emph{high-temperature} bipolaronic
superconductivity is impossible .

However  we have found that small but relatively light  bipolarons
could  exist within the realistic
 range of the \emph{finite-range} EPI with high-frequency optical phonons. The effect appears since  the  finite-range Fr\"ohlich
interaction combined with the long-range Coulomb repulsion  provides
an effective interaction  with a deep attraction minimum for two
holes
 on the neighbouring sites, and repulsive for other hole configurations.   Bipolarons which are both light and small give
  rise to Ogg-Schafroth's bose-condensed state of
charged bosons at  high-temperatures, since the Bose-Einstein
condensate has transition temperature that is inversely proportional
to mass. Our  conclusion is backed up by analytical
\cite{ale5,alekor2} and CTQMC studies \cite{jim2,jim3}. These
studies let us  believe that the following recipes is worth
investigating to look for room-temperature superconductivity
\cite{jim2}: (a) The parent compound should be an ionic insulator
with light ions to form high-frequency optical phonons, (b) The
structure should be quasi two-dimensional to ensure poor screening
of high-frequency phonons polarized  perpendicular to the conducting
planes, (c) A triangular lattice is required in combination with
strong, on-site
 Coulomb repulsion to form the small superlight crab
bipolaron,  (d) Moderate carrier densities are required to keep the
system of small bipolarons close to the dilute regime. Clearly most
of these conditions are already met in the cuprate superconductors.

\section{Acknowledgements}

I would like  to thank   Jim Hague,  Pavel Kornilovitch, and John
Samson  for collaboration and  helpful discussions, and to
acknowledge support of EPSRC (UK) (grant numbers EP/C518365/1 and
EP/D07777X/1).


\begin{thebibliography}{200}
\bibitem{bed} J. G.   Bednorz, K. A. M\"{u}ller,   Z. Phys. B
\textbf{1986}, 64, 189.

\bibitem{bcs} J. Bardeen, L.N. Cooper, and J.R. Schrieffer,
Phys. Rev {\bf 108}, 1175 (1957).

\bibitem{alemot} A. S.  Alexandrov and N. F. Mott, Rep. Prog. Phys. ${\bf 57}$,
1197 (1994).


\bibitem{alebook} A. S. Alexandrov, \emph{Theory of
Superconductivity: From Weak to Strong Coupling} (IoP Publishing,
Bristol, 2003).

\bibitem{edwards}
P. P. Edwards, C. N. R. Rao, N. Kumar, and A. S.  Alexandrov,
ChemPhysChem {\bf 7}, 2015 (2006).

\bibitem{ogg} R. A. Ogg Jr., Phys. Rev. {\bf 69}, 243
(1946).

\bibitem{sha} M. R. Schafroth, Phys. Rev. {\bf 100}, 463 (1955).


\bibitem{blatt}  J. M.  Blatt and S. T. Butler,  Phys. Rev. {\bf %
100}, 476 (1955).

\bibitem{eli} G. M. Eliashberg,  Zh.
Eksp. Teor. Fiz. {\bf 38}, 966 (1960); {\bf 39}, 1437 (1960) [Sov.
Phys. JETP {\bf 11}, 696; {\bf 12}, 1000 (1960)].


\bibitem{ale0} A. S. Alexandrov, Zh. Fiz. Khim.
{\bf 57}, 273 (1983) [Russ. J. Phys. Chem. {\bf 57}, 167 (1983)].

\bibitem{mig}A. B. Migdal, Zh. Eksp. Teor. Fiz. {\bf 34}, 1438
(1958)[Sov. Phys. JETP {\bf 7}, 996 (1958)].

\bibitem{alexandrov:2001}A. S. Alexandrov, Europhys. Lett. {\bf 56}, %
92 (2001).

\bibitem{aleran}
A. S. Alexandrov  and J. Ranninger, Phys. Rev. B {\bf 23} 1796
(1981), ibid {\bf 24}, 1164 (1981).


\bibitem{mel} E. V. L.  de Mello and J.
Ranninger, Phys. Rev. B {\bf 58}, 9098 (1998).

\bibitem{and2} P. W. Anderson,  \emph{The Theory of Superconductivity in the
Cuprates},  Princeton Univ. Press, Princeton NY (1997).


\bibitem{aub93} S. Aubry,
 in {\em Polarons and Bipolarons in High Tc Superconductors and related
 materials},
eds.  E. K. H. Salje, A. S. Alexandrov and W. Y. Liang (Cambridge
University Press, Cambridge, 1995), 271.

\bibitem{alexandrov:1991}
A. S. Alexandrov, Physica C (Amsterdam)
  {\bf 182}, 327 (1991).

\bibitem{alegap}  A. S. Alexandrov, J. Low Temp. Phys. {\bf %
87}, 721 (1992);   A. S. Alexandrov and N. F. Mott, J.
Superconductivity: Incorporating Novel Magnetism, {\bf 7}, 599
(1994).

\bibitem{trugman} J. Bon\v{c}a, T.~Katrasnic, and
S.~A. ~Trugman, Phys. Rev. Lett. {\bf 84}, 3153 (2000).

\bibitem{lamagna} A. La Magna and R. Pucci, Phys. Rev. B
{\bf 55}, 14886 (1997).

\bibitem{Macridin}
A. Macridin, G.~A. Sawatzky, and M. Jarrell, Phys. Rev. B
 {\bf 69}, 245111 (2004).

\bibitem{micnas1981} S. Robaszkiewicz, R.~Micnas, and K.~A. Chao, Phys.~Rev.~B {\bf 23}, 1447 (1981).





\bibitem{ale5} A. S. Alexandrov, Phys. Rev. B {\bf 53}, 2863
(1996).

\bibitem{alekor}
A. S. Alexandrov and P. E. Kornilovitch,  Phys. Rev. Lett. ${\bf
82}$, 807 (1999).

\bibitem{spencer}  P. E. Spencer, J. H. Samson,  P. E. Kornilovitch, and A. S.
Alexandrov, Phys. Rev. B {\bf 71}, 184319 (2005).

\bibitem{jim}  J. P. Hague, P. E. Kornilovitch, A. S. Alexandrov, and J. H. Samson,
Phys. Rev. B {\bf 73}, 054303 (2006).



\bibitem{alekor2}  A. S. Alexandrov and P. E. Kornilovitch, J.
Phys.: Condens. Matter {\bf 14}, 5337 (2002).


\bibitem{jim2} J. P. Hague, P. E. Kornilovitch,
J. H. Samson, and A. S. Alexandrov, Phys. Rev. Lett. {\bf 98},
037002 (2007).

\bibitem{jim3} J. P. Hague, P. E. Kornilovitch,
J. H. Samson, and A. S. Alexandrov, submitted to the special Mott's
issue of J. Phys.: Condens. Matter (2007).

\bibitem{lan} I. G. Lang and  Y.~A.~Firsov, Zh. Eksp. Teor. Fiz.
{\bf 43},  1843 (1962) [Sov. Phys. JETP {\bf 16}, 1301 (1962)].


\bibitem{ale2} A. S. Alexandrov, Phys. Rev. B {\bf 46}, 2838 (1992).

\bibitem{plasphon}
A. S. Alexandrov, Sol St. Commun.
 {\bf 81}, {965} (1992).

 \bibitem{splitting} H. Rietschel, L. Pintschovius, and W. Reichardt, \bibinfo{year}{1989}, Physica C (Amsterdam)  {\bf 162}, 1705
 (1989).

\bibitem{cat} C. R. A. Catlow, M. S. Islam,  and X. Zhang,  J.
Phys.: Condens. Matter {\bf 10}, L49 (1998).

\bibitem{aledwave} A. S. Alexandrov,  J.  Superconductivity: Incorporating Novel Magnetism, {\bf 17}, 53
(2004).

\bibitem{ver} G. Verbist, F. M. Peeters, and J. T. Devreese, Phys.
Rev. B {\bf 43}, 2712 (1991); Solid State Commun. \textbf{76}, 1005
(1990).








\end{thebibliography}
\end{document}